# Estimating the unconfined compressive strength of carbonate rocks using gene expression programming

**Saeid R. Dindarloo***

Department of Mining and Nuclear Engineering, *Missouri University of Science and Technology, Rolla, MO, USA.*
*Email:srd5zb@mst.edu

**Elnaz Siami-Irdemoosa**

Department of Geosciences and Geological and Petroleum Engineering, Missouri University of Science and Technology, Rolla, MO, USA.

**Abstract**

Conventionally, many researchers have used both regression and black box techniques to estimate the unconfined compressive strength (UCS) of different rocks. The advantage of the regression approach is that it can be used to render a functional relationship between the predictive rock indices and its UCS. The advantage of the black box techniques is in rendering more accurate predictions. Gene expression programming (GEP) is proposed, in this study, as a robust mathematical alternative for predicting the UCS of Iran's carbonate rocks. The two parameters of total porosity and P-wave speed were selected as predictive indices. The proposed GEP model had the advantage of the both traditionally used approaches by proposing a mathematical model, similar to a regression, while keeping the prediction errors as low as the black box methods. The GEP outperformed both artificial neural networks and support vector machines in terms of yielding more accurate estimates of UCS. Both the porosity and the P-wave velocity were sufficient predictive indices for estimating the UCS of the carbonate rocks in this study. Nearly, 95% of the observed variation in the UCS values was explained by these two parameters (i.e., $R^2$ =95%).

1. **Introduction**

The unconfined compressive strength (UCS) is a rock strength index that is widely used in geothechnical, mining, and construction projects. For estimating/measuring the UCS, two methods can be used: (i) direct UCS measurement, and (ii) indirect UCS estimation. Both the ASTM [1] and the ISRM [2] have specific laboratory and in situ testing procedures for the direct measurement of UCS. Sophisticated representative samples must however be prepared before direct tests can be conducted. These samples are, at times, difficult to obtain. For example, the preparation of standard specimens in highly weathered/fractured rocks is both a challenging and time/money consuming task. Thus, a considerable amount of research has been devoted to the indirect UCS estimation from a range of mechanical, physical, and petrological indices in an effort to address these limitations. The indirect UCS estimation methods can be categorized into two classes: black box [3-6] and regression approaches [7-10]. In the black box approach, that is mainly applied by soft computing techniques such as artificial neural networks (ANNs), a kernel function maps a vector of input (independent) variables into a vector of response (dependent) variables. After learning from a portion of the available datasets (training dataset), the model will be able to predict values of the response vector for the other portion (testing dataset). These techniques, however, do not render a functional relationship between the input and output vectors. Thus, they are referred to as black box techniques. In contrast, the regression methods (e.g., multiple variable and support vector regression) propose a functional relationship between the variables. This results in a better understanding of both the underlying process and interactions between variables. A number of studies have found that the soft computing methods perform better than the regression approach performs in terms of accurate predictions [11-13]. Thus, the advantage of the black box approach is in proposing better estimates while the advantage of the regression is in proposing a functional relationship between the variables. The genetic programming (GP) technique was adapted in this study that combines the individual advantages of other methods into a single framework. The GP simultaneously reduces the prediction errors (comparable to the black box techniques) and renders a functional relationship (comparable to the regression approach).

Genetic programming is among the recent advances made in mathematical modeling, which are based on evolutionary principles. GP is a development of the genetic algorithm (GA) that generates thousands of computer programs to solve regression and classification problems. Unlike the conventional GA technique that seeks optimal values for some predefined parameters, GP manipulates many programs, based on the Darwinian evolution theory, to find both the best models and the best parameters for a given set of variables. Introduced by Ferreira [14], gene expression programming (GEP) was developed from the original GP. The entities in GEP (unlike those in GP) are in the form of the same length strings [15].

Although the technique emerged only recently, numerous applications have been discussed. GP has been applied to a range of geological, environmental, and civil problems [16-20]. The GEP method has not, however, been used to predict the UCS of carbonate rocks. Unlike other soft computing techniques, such as ANNs, GEP renders a (none) linear relationship between the independent variables and the response. Hence, this technique helps with providing a better understanding of the phenomena. In order to demonstrate the superiority of the GP in predicting the UCS, the technique is applied in the case of carbonate rocks. Moreover, the results are compared with two techniques of SVR (regression approach) and ANNs (black box approach).

The GEP was used in this study to construct a functional relationship between P-wave velocity and total porosity (independent variables) and UCS (dependent variables) of 117 samples of carbonate rocks. The results were compared with support vector machine (SVR) and artificial neural networks (ANNs) in terms of modeling capability and prediction accuracy.

2. **Materials and Methods**
2.1 **Dataset**

Density [21], impact strength [22], porosity [23-24], and fabric [25] are among indices most widely used to predict UCS. We used in this study: P-wave velocity and total porosity to estimate UCS [4, 26]. A carbonate formation of sedimentary limestones in a dam site located in southern Iran was investigated. A total of 117 representative core specimens were prepared for the measurement of UCS (per ASTM) [1]. The height-to-diameter ratio was approximately 2.7, and the 2 ends were flattened. The net total porosity was calculated as in Eq. (1):

$$n = 1 - \rho_d / \rho_s \quad (1)$$

In Eq. (1):  n, $\rho_d$ , and $\rho_s$ are porosity, dry density, and density of the solid part, respectively.

Furthermore, the P-wave velocity within dry samples was measured per the method suggested by the Ref. [2]. Two-thirds of the data set was used to build the model. The remaining 39 (one-third) samples were used to test the model. The randomly selected test dataset is presented in Table 1.

Table 1 The randomly selected testing dataset

| Sample | n (%) | v (m/s) | UCS (Mpa) |
| --- | --- | --- | --- |
| 1 | 24.9 | 2551.2 | 69.1 |
| 2 | 18.9 | 2766 | 66.5 |
| 3 | 23 | 2618.4 | 65.6 |
| 4 | 19.8 | 2900 | 64 |
| 5 | 19.7 | 2679 | 63.6 |
| 6 | 29.3 | 3135 | 60.9 |
| 7 | 33.3 | 2775 | 69.5 |
| 8 | 35.5 | 2903.3 | 71.8 |
| 9 | 31.8 | 3341.1 | 59.1 |
| 10 | 24.2 | 3033.1 | 57.2 |
| 11 | 14.1 | 3016.5 | 56.2 |
| 12 | 20.6 | 3107.8 | 53.8 |
| 13 | 35.3 | 3832.7 | 61.3 |
| 14 | 28.5 | 3560.8 | 58 |
| 15 | 18.8 | 3552.7 | 50.9 |
| 16 | 31.5 | 3884.7 | 50.8 |
| 17 | 18 | 3973.4 | 46.6 |
| 18 | 25.7 | 4297.6 | 44.5 |
| 19 | 15.9 | 4014.4 | 44.4 |
| 20 | 31.7 | 4390.1 | 54.2 |
| 21 | 29.4 | 4392.8 | 53.9 |
| 22 | 21.5 | 4383.5 | 43.9 |
| 23 | 18.5 | 4330.2 | 42.7 |
| 24 | 18.3 | 4518.2 | 41.8 |
| 25 | 22.2 | 4548.3 | 41.4 |
| 26 | 20.4 | 4651.9 | 40.9 |
| 27 | 24.9 | 4395.8 | 50.6 |
| 28 | 14.2 | 4240.8 | 40.4 |
| 29 | 34.6 | 4822 | 48.8 |
| 30 | 14.9 | 4586.4 | 38.2 |
| 31 | 12.6 | 4870.8 | 36 |
| 32 | 20 | 4990.9 | 35.5 |
| 33 | 9.9 | 5030.4 | 35.4 |
| 34 | 23.2 | 5047 | 44.8 |
| 35 | 10.1 | 4532 | 34.4 |
| 36 | 14.1 | 5580.5 | 32.8 |
| 37 | 17.3 | 5051.6 | 36.6 |
| 38 | 15.3 | 5217.8 | 34.6 |
| 39 | 11.2 | 4946.1 | 34.5 |

**2.2 Genetic Programming**

Koza (1992) developed GP as an extension of the traditional GA. In GP the adaptation procedure is composed of thousands of computer programs with different sizes and structures [27]. The optimization task in GA is to find (near) optimal values for a set of given variables. In GP, however, both the solution's structure (e.g., type of the fitness function for a regression problem) and the optimal values of its associated parameters are derived together. In GP, thousands of solutions (computer programs) are generated and evolved consecutively according to the Darwinian principle of survival. A search for the solution begins with a population of completely random programs (solutions) generated from a predefined set of available functions (e.g., arithmetic functions) and terminals (independent variables). All programs are measured against a fitness function (e.g., root mean square error within a regression problem). Consequently, only the best programs survive and will breed to the next generation. The GP can be represented as a hierarchically structured tree comprised of functions and terminals. A simple representation of a GP tree for function $y = Z^2(\sin x + C_1)$ is illustrated in Fig. 1. The tree reads from left to right and bottom to top. The fittest solutions (smallest error) will be chosen to generate a population of new offspring programs for the next generation, mimicking the Darwinian principle of survival. Next, several genetic operations, namely, mutation, inversion, transposition, recombination and crossover, will generate new offspring from the previous generation's fittest programs. In the mutation, as the most important genetic operator, the operator selects a random node and replaces it with a new node or subtree. Either the error or fitness function is used to evaluate the new offspring. This process continues until a predefined threshold is reached in terms of the best fit or error.

The result of application of GP in modeling UCS, using P-wave velocity and porosity, are presented and compared with SVR and ANNs in Sec. 3.

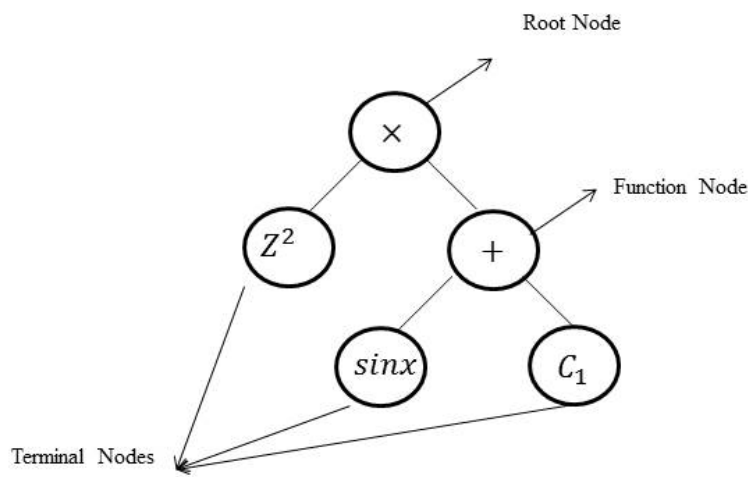

Fig. 1 GP tree representation of $Z^2(\sin x + C_1)$

### 3. Results and Discussion

#### 3.1 Genetic programming

Gene expression programming regression of the training dataset produced the nonlinear equation in Eq. (2). The results of the model for predicting the testing dataset (not used for model building) are illustrated in the scattergram of Fig. 2.

$$\text{UCS}_{(\text{MPa})} = -\frac{1}{88}\frac{n^3 + v}{1 + n^2} + \frac{v(n-2)}{4+v} + \frac{12 \times 10^4}{v} \quad (2)$$

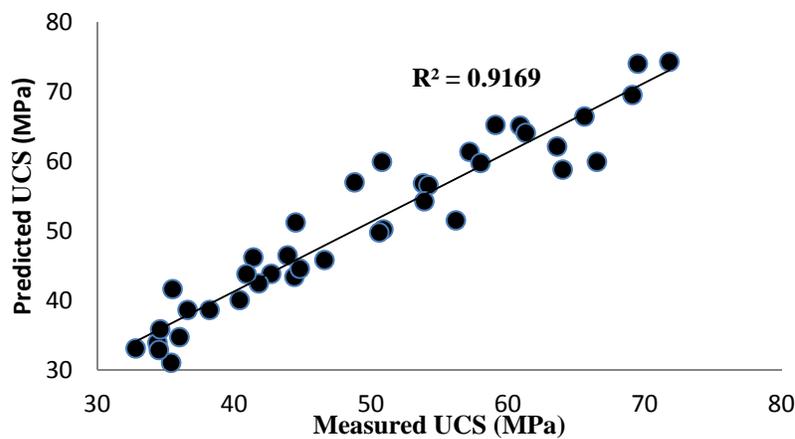

Fig. 2 Scattergarm of measured vs. GEP predicted UCS (testing dataset)

The mean absolute percentage error (MAPE in Eq. 3) was calculated as 5.7% for the testing dataset.

$$MAPE = \frac{1}{N}\sum_{i=1}^{N}\left|\frac{y_{meas} - y_{pred}}{y_{meas}}\right| \times 100 \quad (3)$$

where

$y_{meas}$ and $y_{pred}$ are the measured and predicted UCS values, respectively.

#### 3.2 Support Vector Machine

Support vector machine (SVM) is a popular machine learning method used for both classification (SVC) and regression (SVR). SVR was used in this study for the purpose of predicting the UCS from n and v. LIBSVM is a library of SVM algorithms [31] that was used for the SVR modeling in this study. The results of the SVR regression are plotted on the scattergram illustrated in Fig. 3.

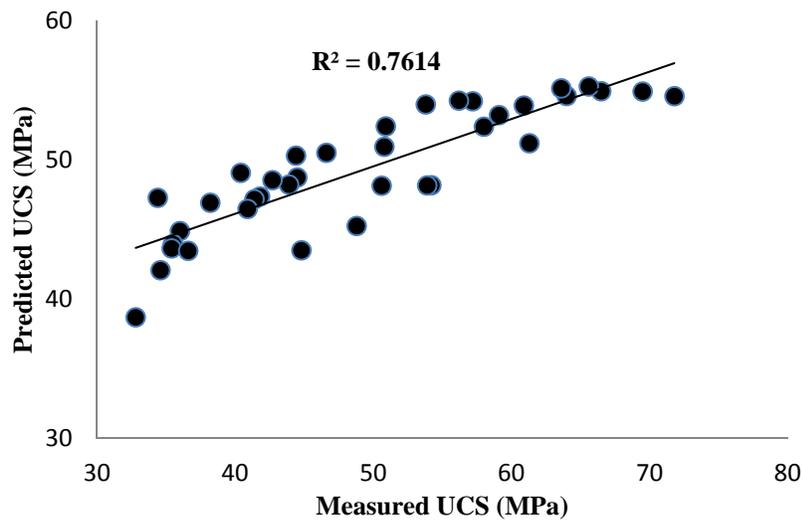

Fig. 3 Scattergarm of measured vs. SVM predicted UCS (testing dataset)

The MAPE value for the SVM model was calculated as 13.8 %.

### 3.3 Neural Networks

A feed-forward back propagation neural network architecture, with one hidden layer and three neurons in the hidden layer, was constructed and trained with a rate and momentum of 0.3 and 0.2, respectively. The results of the ANNs regression are plotted on the scattergram depicted in Fig. 4

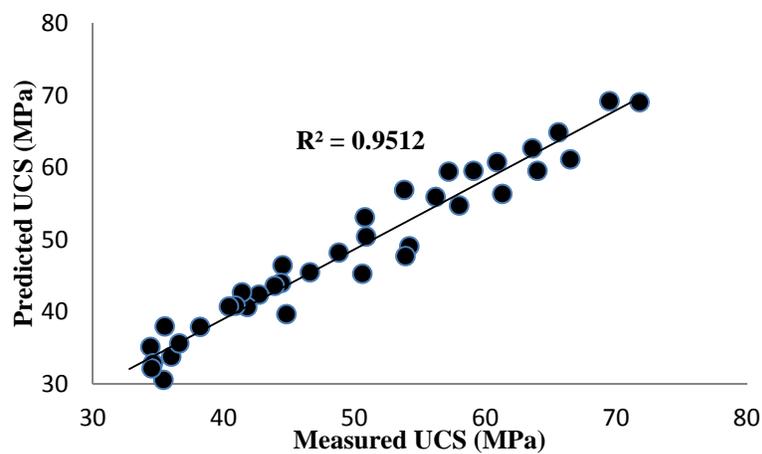

Fig. 4 Scattergarm of measured vs. ANNs predicted UCS (testing dataset)

The MAPE value for the ANNs model was calculated as 4.4 %.

The three proposed models were compared to one another in terms of the coefficient of determination and MAPE (see Table 2).

Table 2 Model comparison

|  | SVM | GEP | ANN |
|---|---|---|---|
| MAPE (%) | 13.8 | 5.9 | 4.4 |
| $R^2$ (%) | 76 | 92 | 95 |

Both the ANNs and the GEP outperformed the SVM with more accurate predictions. The $R^2$ and MAPE values of GEP were very close to those obtained by the ANNs. The GEP proposed a compromised solution between the regression (SVM) and the black box (ANNs) approaches. Although the predictions of GEP are negligibly worse than the ANNs, but a functional relationship (Eq. 2) is proposed. This relationship can be employed in similar situations promptly.

4. Conclusions

The application of genetic programming as a robust mathematical tool for an indirect prediction of the UCS of carbonate rocks was demonstrated. Although many studies have utilized other soft computing tools (particularly, ANNs) the GP demonstrated to be a promising alternative. Genetic programing solved the problem of conventional regression algorithms by proposing considerably more accurate predictions. It also solved the shortcomings of the conventional black box modeling techniques by proposing a mathematical model through the proposed functional relationship for UCS.

Estimating the unconfined compressive strength of carbonate rocks using gene expression programming    Dindarloo & Siami-Irdemoosa (2015)